\documentclass[twocolumn,pra,longbibliography]{revtex4-1}
\usepackage{graphicx}
\usepackage{float}
\usepackage{dcolumn}
\usepackage{bm}
\usepackage{color}
\usepackage{amsmath}
\usepackage{amssymb}
\usepackage{amsfonts}
\usepackage{esint}
\usepackage{times}
\usepackage{xcolor}
\usepackage{phaistos}
\usepackage{braket}
\usepackage{comment}

\newcommand{\ou}{%
  \mathrel{%
    \vcenter{\offinterlineskip
      \ialign{##\cr$\,<$\cr\noalign{\kern-0.6pt}$(>)$\cr}%
    }%
  }%
}

\begin{document}

\title{Integration of the Berry curvature on a qubit state manifold by coupling to a quantum meter system}

\author{Peng Xu$^{abcd}$}\email{pengxu@njupt.edu.cn}
\author{Shi-Liang Zhu$^{de}$}
\author{Klaus M{\o}lmer$^b$}\email{moelmer@phys.au.dk}
\author{Alexander Holm Kiilerich$^{b}$}
\affiliation{$^a$ Institute of Quantum Information and Technology, Nanjing University of Posts and Telecommunications, Nanjing, Jiangsu 210003, China \\
$^b$ Department of Physics and Astronomy, Aarhus University, 8000 Aarhus C, Denmark \\
$^c$ State Key Laboratory of Quantum Optics and Devices, Shanxi University, Taiyuan, 030006, China \\
$^d$ National Laboratory of Solid State Microstructures, Nanjing University, Nanjing 210093, China\\
$^e$ Guangdong Provincial Key Laboratory of Quantum Engineering and Quantum Materials,
GPETR Center for Quantum Precision Measurement,  Frontier Research Institute for Physics, SPTE,
South China Normal University, Guangzhou 510006, China}

\begin{abstract} 
We present a scheme that allows integration of the Berry curvature and thus determination of the Chern number of a qubit eigenstate manifold. Our proposal continuously couples the qubit with a meter system while it explores a quasi-adiabatic path in the manifold. The accumulated change of one of the meter observables then provides an estimate of the Chern number. By varying the initial state of the meter, we explore the delicate interplay between the measurement precision and the disturbance of the qubit. A simple argument yields a correction factor that allows estimation of the Chern number, even when the qubit is significantly disturbed during the probing. 
The Chern number arises from the geometric phase accumulated during the exploration, while we observe the dynamic phase to produce a broadening of the meter wave function. We show that a protocol, relying on three subsequent explorations, allows cancellation of the dynamic phase while the geometric phase is retained.
\end{abstract}

\maketitle

\section{Introduction}
Since the possibility of topological phases of matter in one and two dimensions was first realized \cite{Kosterlitz1973, Thouless1982, Haldane1983} and observed in condensed-matter systems \cite{Klitzing1980, Tsui1982}, they have been subject to ongoing research \cite{Bernevig2006, Hasan2010, Moore2010} and proposals for implementation in different systems \cite{Carusotto2019, Dalibard2019,DanWei2018}. The Berry phase \cite{Berry1985} is associated with the evolution of a quantum system under adiabatic variation of the Hamiltonian and gives rise to topological invariants that account for how the eigenstates of a continuously varied Hamiltonian with a closed  manifold parameter space are connected. The topological Chern number is such an invariant, and its restriction to integer values explains the robustness of physical phenomena such as the integer quantum Hall effect \cite{Klitzing1986, Haldane1988, Zhang2005}. This robustness implies reduced sensitivity to small perturbations, suggesting quantum systems with nontrivial topology as promising platforms for quantum computing \cite{Nayak2008}.

Because of its dependence on the eigenstate of a manifold of different Hamiltonians, we cannot measure the Chern number as a single quantum mechanical observable. However, it has been noted recently that the slowly quenched dynamics of a quantum system permits estimation of the Berry curvature as the expectation value of the gradient of the Hamiltonian \cite{Gritsev2012}. Hence, the Chern number can be determined in an experiment by measuring the corresponding physical observable; either in a sequence of projective measurements performed after sweeps of the Hamiltonian towards different points in the manifold \cite{Schroer2014,Roushan2014}, or by accumulation of weak monitoring signals during continuous sweeps across the parameter manifold \cite{Xu2017}. The former approach requires a large number of repeated experiments, while to avoid perturbing the later state of the system by the earlier measurements, the latter permits only weak and hence noisy measurements .   

\begin{figure*}
\begin{center} 
\includegraphics [scale=0.28] {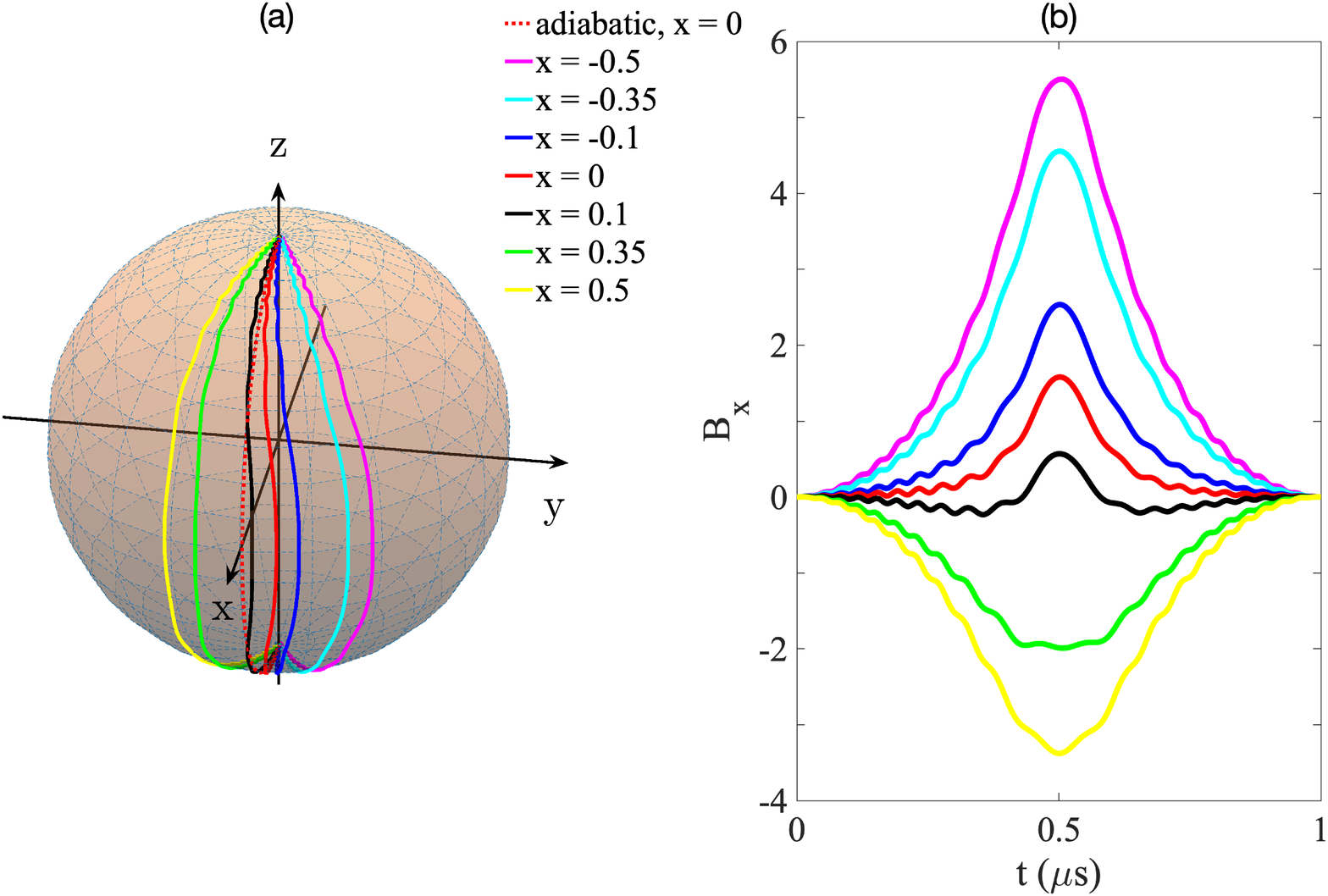}
\includegraphics [scale=0.3] {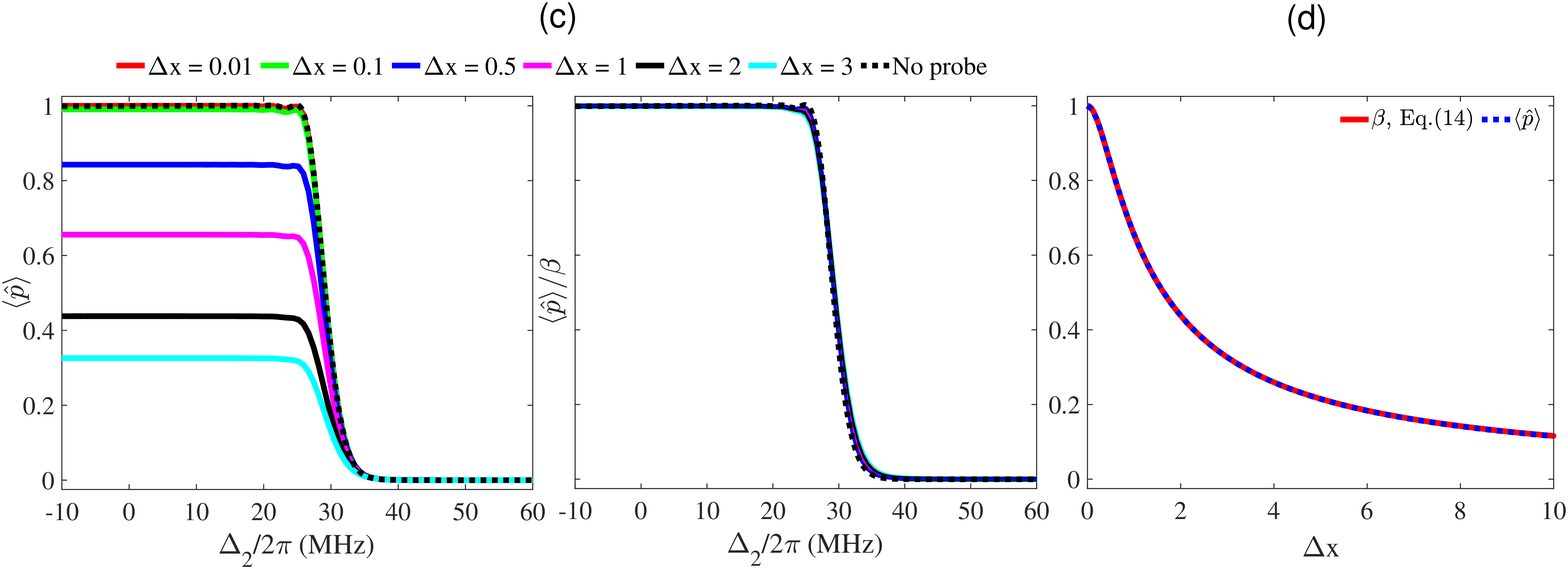}
\end{center}
\caption{Influence of the interaction with the meter system on the qubit dynamics, the Berry curvature and the Chern number. Results are shown for $t_q = 1\mu s$, $\Delta_1 = 2\pi\times 30$MHz, $\Omega_1 = 2\pi\times 10$MHz and $\Delta_2 = 2\pi\times 0.3$ MHz in (a) and (b). 
(a) Bloch sphere representation of the qubit evolution for different values of $x \in (-0.5, -0.35, -0.1, 0, 0.1, 0.35, 0.5)$ from right to left at $t = 0.5\mu s$. The dotted red line tracks the fully adiabatic evolution for $x = 0$. 
(b) The corresponding $x$-dependent Berry curvature Eq.~(\ref{eq:Bx}) as a function of time shown for $x \in (-0.5, -0.35, -0.1, 0, 0.1, 0.35, 0.5)$ from above.
(c) Left panel: The expectation value $\langle \hat{p}\rangle$ of the meter momentum observable at the final time $t_q$, estimating the true Chern number indicated by the dotted black curve.
Results are shown as a function of $\Delta_2$ and for different values of $\Delta x \in (0.01, 0.1, 0.5, 1, 2, 3)$ from above at $\Delta_2/2\pi = 0$ MHz. Right panel: The corrected estimator of the Chern number $\langle \hat{p}\rangle/\beta$.
(d) The expected meter momentum at the final time $\langle \hat{p}\rangle$ for $\Delta_2 = -2\pi\times 10$MHz is shown along with the analytical expression Eq.~(\ref{beta}) as a function of $\Delta x$.   
\label{F1}}
\end{figure*}

In this work, we propose a theoretical scheme to measure the Chern number by coupling the quantum system of interest to an ancillary meter system while its Hamiltonian is varied across the parameter range of interest. We imagine a meter system with canonical position and momentum variables $x$ and $p$, and introduce an interaction Hamiltonian $\hat{H}_I\propto\hat{x}\hat{A}$, where $\hat{A}$ can represent any system observable. Assuming that $\hat{x}$ is a quantum nondemolition (QND) observable of the meter system, the interaction causes an accumulated change of $\hat{p}$ in the Heisenberg picture which is nothing but the temporal integral of $A$. Hence, if $\hat{A}$ is chosen proportional to the system observable that yields the Berry curvature as the system Hamiltonian is varied, a measurement of the change in $\hat{p}$ provides a good estimation of Chern number.

The article is outline as follows. In Sec.~\ref{sec:system}, we first discuss a two-level Rabi model which explores the topology of a slowly varying Hamiltonian.
We then proceed to introduce our measurement protocol, relying on a suitable coupling of the two-level system to a measurement meter.
In Sec.~\ref{sec:Results}, we investigate the performance of our scheme by numerical simulations and analytic arguments. Subsequently we propose a multi quench protocol which ensures a refocusing of the meter wave function, thereby allowing a high precision read-out of the Chern number. 
In Sec.~\ref{sec:Conclusion}, we conclude and provide an outlook.

\section{A topological spin system}\label{sec:system}

We illustrate our proposal by an analysis of the same system as studied in \cite{Schroer2014,Roushan2014} and \cite{Xu2017}; i.e., a quantum two-level system subject to driving by a detuned electromagnetic field. The detuning and Rabi frequency explore a range of parameters, and the model is equivalent to a spin $1/2$ particle subject to a magnetic field. The surface explored by the effective magnetic field vector plays a role similar to the Brillouin zone explored by a solid state system in a given Bloch band, and hence permits the simulation of topological bands by the study of the evolution of a single two-level system. 

\subsection{Rabi model}\label{sec:topoModel}
Following \cite{Schroer2014,Xu2017}, we consider a quantum system with two levels $|g\rangle$, $|e\rangle$ and transition frequency $\omega_q$ driven by a field of frequency $\omega_d$.
In the rotating frame at the drive frequency $\omega_d$, the Hamiltonian may be written ($\hbar = 1$)
\begin{equation}\label{eq:H} 
\hat{H}_q = \frac{1}{2} [\Delta\hat{\sigma}_z + \Omega\hat{\sigma}_x \cos(\phi) + \Omega\hat{\sigma}_y \sin(\phi)].
\end{equation}  
Here $\hat{\sigma}_x$, $\hat{\sigma}_y$ and $\hat{\sigma}_z$ are the Pauli operators, $\Delta = \omega_q - \omega_d$ is the system-field detuning, the Rabi frequency $\Omega$ denotes the product of the field amplitude and the transition dipole moment, and $\phi$ is the relative phase of the field. The experiment reported in Ref.~\cite{Schroer2014} concerns a superconducting qubit driven by a microwave field and for concreteness we shall have the parameters of this setup in mind. However, the Hamiltonian~(\ref{eq:H}) may also be realized by optical driving of an atomic system or by exposing a spin-$1/2$ particle to a magnetic field with components $(B_x,B_y,B_z) \propto (\Omega\cos\phi,\Omega\sin\phi,\Delta)$.

A dynamical quench is performed by slowly changing the detuning and Rabi frequency according to
\begin{align}\label{eq:quench}
\begin{split}
\Delta &= \Delta_{1} \cos\theta + \Delta_{2},
\\
\Omega &= \Omega_{1} \sin\theta,  
\end{split}
\end{align}
where the quench parameter $\theta$ is changed linearly with time, $\theta(t) = \nu t$ at a speed, $\nu = \pi/t_{q}$ determined by the total quench time $t_{q}$. 
The relative values of $\Delta_{1}$ and $\Delta_{2}$ dictate the resulting evolution of the two level system. If
$|\Delta_{2}|/|\Delta_{1}|<1$, the microwave frequency performs a chirp across the qubit resonance and if $t_q$ is sufficiently long,
the qubit Bloch vector passes from the north (or south) to the south (north) along the adiabatic eigenstate with Bloch vector representation
\begin{align}\label{bloch AD}
\vec{R} 
= \frac{1}{\sqrt{\Omega^2+\Delta^2}}
\begin{pmatrix}
\Omega\cos \phi \\ \Omega\sin \phi\\\Delta.
\end{pmatrix}
\end{align}
However, if $|\Delta_{2}|/|\Delta_{1}|>1$, the initial and final states of the adiabatic evolution are represented by the same state on the Bloch sphere. 

If we fix $\phi = 0$ and initialize the qubit in the ground state $\theta(t=0) = 0$, the Berry curvature associated with this quench is given by \cite{Xu2017} 
\begin{align}\label{eq:Berry_y}
B(\theta) = \frac{\Omega_{1}}{2\nu}\langle \hat{\sigma}_y\rangle \sin \theta.
\end{align}
The adiabatic Bloch vector evolution occurs in the $(\hat{\sigma}_x,\hat{\sigma}_z)$-plane, and values $\braket{\hat{\sigma}_y}\neq 0$ arise as non-adiabatic corrections.
Due to the azimuthal symmetry of the Bloch sphere dynamics, we do not need to probe the entire sphere and the topological Chern number has the value \cite{Xu2017}
\begin{equation}\label{Ctwo_level} 
C = -\int_{0}^{\pi} B(\theta) d\theta.
\end{equation} 
We have simulated the dynamics with a finite duration of the quench and evaluated this integral; see the dotted line in Fig.~\ref{F1}(c), which takes the value $1$ if $|\Delta_{2}|/|\Delta_{1}|<1$ (the qubit state flips from $\ket{e}$ to $\ket{g}$ during the quench) and the value $0$ if $|\Delta_{2}|/|\Delta_{1}|>1$ (the qubit state remains $\ket{e}$ after the quench). For values around $|\Delta_{2}| = |\Delta_{1}|$, longer quench times are necessary to reveal the steep transition between $C=1$ and $C=0$. 

\subsection{Measurement protocol}
In Ref.~\cite{Schroer2014} the Chern number was determined experimentally by projectively measuring the qubit $\hat{\sigma}_y$-component after a large number of partial quenches towards different values of the angle argument $\theta\in [0,\pi]$. As an alternative to partial sweeps followed by destructive measurements, we proposed in Ref.~\cite{Xu2017} to perform a continuous, weak measurement of the qubit $\hat{\sigma}_y$-component during a full quench. By applying a measurement controlled feedback we show that it is possible to partially counteract the measurement back action on the qubit system and recover the Chern number with only few repetitions of the experiment. 

Here we propose to employ an ancillary quantum system with continuous variables $x$ and $p$, to accumulate the Berry curvature contributions to the Chern number during a single sweep of the argument $\theta\in [0,\pi]$. Depending on the specific settings, this model may be implemented in different ways. A qubit may, for instance, couple to a quantized electromagnetic or acoustic mode which act as the measurement meter. Similarly, the internal discrete states in a trapped ion or atom may couple to its center of mass motion. 

In a suitable rotating frame, the oscillator Hamiltonian system vanishes, and we assume the coupling
\begin{equation}\label{Hi}
\hat{H}_{I} = -g(t)\hat{\sigma}_y\hat{x},
\end{equation} 
where $g(t)$ is a controllable coupling strength. 
The interaction resembles the Stern-Gerlach splitting of an atomic wave packet in a static magnetic field whose strength is slowly varied according to $g(t)$ \cite{Gerlach1922}. However, it is desirable to implement the model in a more controllable quantum optical setting.
Although the coupling~(\ref{Hi}) is not realized in the Jaynes-Cummings model (JCM) of conventional QED setups where the counter rotating terms are suppressed, it is possible to recover these by combining the JCM with the so-called anti-JCM. This may be achieved in a Raman scheme by driving both the red and blue sidebands of the Raman transition \cite{Meekhof1996}.

In the Heisenberg picture, the equation of motion for the (generalized) momentum operator is given by 
\begin{equation}\label{dp}
\frac{\mathrm{d} \hat{p}}{\mathrm{d} t} = -g(t)\hat{\sigma}_y.
\end{equation}  
By controlling the coupling $g(t)$, the momentum observable of the measurement meter thus records a weighted integral of the qubit $\hat{\sigma}_y$ observable. In particular, we propose to set $g(t) = \Omega_1\frac{\sin\theta}{2}$ such that $\frac{\mathrm{d} \hat{p}}{\mathrm{d} t}$ follows the Berry curvature, allowing the total change in momentum to precisely yield the Chern number,
\begin{equation}\label{pt}
\langle \hat{p}(t_q)\rangle-\langle \hat{p}(0)\rangle = -\Omega_1 \int \frac{\sin\theta}{2} \langle \hat{\sigma}_y \rangle d\theta.
\end{equation}  
Note that a measurement of the momentum yields a random outcome and that the Chern number is thus determined as the mean value of many such measurements. To address the practical prospects of the method, we shall proceed to study the evolution of the qubit and oscillator state in more detail. 

\section{Results and Discussion}\label{sec:Results}

For concreteness, we assume that the meter is initially prepared in a pure Gaussian state with vanishing mean position and momentum, 
\begin{equation}\label{eq:phiX}
\varphi(x) = \frac{1}{\sqrt[4]{2\pi(\Delta x)^2}}e^{-\frac{x^{2}}{4(\Delta x)^{2}}},
\end{equation}
for which the uncertainties in position and momentum fulfill $\Delta x \Delta p = 1/2$. For $\Delta x  = 1/\sqrt{2}$, the position and momentum variances are equal, while for $\Delta x \gtrless 1/\sqrt{2}$, the meter is prepared in a squeezed state with a smaller or larger momentum uncertainty $\Delta p = 1/(2\Delta x)$.

Since the meter position $\hat{x}$ is conserved under the Hamiltonian $\hat{H}_q + \hat{H}_I$, the qubit evolution may be solved independently for each value of $x$, yielding a set of $x$-dependent, pure-state trajectories for the qubit state 
\begin{align}\label{eq:wf}
\ket{\chi(x,t)} = c_e(x,t)\ket{e} + c_g(x,t)\ket{g},
\end{align}
as shown for a few values of $x$ in Fig.~\ref{F1}(a). The interaction with the meter state causes each trajectory to deviate substantially from the bare qubit case (x = 0) and causes a rotation towards $\pm y$ according to the sign of $x$.
The full qubit-meter state is obtained by weighing the $x$ trajectories by the initial meter state in a superposition 
\begin{align}\label{eq:fullState}
\ket{\Psi(t)}  = \int \varphi(x) \ket{\chi(x,t)}\otimes \ket{x} dx.
\end{align}

While one might expect that a meter state with reduced momentum uncertainty would benefit the measurement of the Chern number, we note that the large $x$ values explored by the interaction Hamiltonian lead to correspondingly stronger perturbations of the qubit system and may hence significantly alter the qubit dynamics (see the Bloch vectory trajectories for different $x$ in Fig.~\ref{F1}(a)). This impacts the value of $\braket{\hat{\sigma}_y}$ and thereby the evolution of the momentum of the meter. The resulting influence on our estimate from $\ket{\chi(x,t)}$ of the Berry curvature
\begin{align}\label{eq:Bx}
B_x(\theta) = \frac{\Omega_{1}}{2\nu}\langle \chi(x,t)|\hat{\sigma}_y|\chi(x,t)\rangle \sin \theta, 
\end{align}
as defined in Eq.~\eqref{eq:Berry_y}, is depicted for different $x$ in Fig.~\ref{F1}(b). 
The left panel in Fig.~\ref{F1}(c) shows the candidate value of the Chern number, extracted from the expectation value of the meter momentum $\braket{\hat{p}}$ in the state~(\ref{eq:fullState}) at the final time $t = t_q$ for different widths $\Delta x$ of the initial meter state.
All curves show a transition between a vanishing and a non-vanishing value around $\Delta_2 = \Delta_1$ ($\Delta_1 = 3\Omega_1 = 2\pi\times 30 $MHz) and are flat beyond the transition regime.
Even so, we observe that for larger values of $\Delta x$, the Chern number estimate by $\braket{\hat{p}(t_q)}$ to an increasing degree deviates from the integer value $C=1$  associated with the topological properties of the qubit model for $\Delta_2 < \Delta_1$, as discussed in Sec.~\ref{sec:topoModel}.

At a first glance this seems to limit the scope of our protocol to narrow initial states with $\Delta x \ll 0.1$, resulting, unfortunately, in a large momentum uncertainty such that many experiments are required to recover the momentum change with sufficient precision. Upon a closer look, however, we note that in Fig.~\ref{F1}(a) the different Bloch sphere trajectories seem to be merely rotated versions of each other, and hence there may be a simple relationship between their candidate time dependent values of $\braket{\hat{\sigma}_y}$.   This can be further quantified by rewriting the full Hamiltonian (for $\phi = 0$) in the following form
\begin{align}\label{eq:Hfull}
\hat{H}_q + \hat{H}_I = \frac{1}{2} [\Delta\hat{\sigma}_z + \tilde{\Omega}\hat{\sigma}_\xi],
\end{align}
with $\hat{\sigma}_\xi = \cos{(\xi)}\hat{\sigma}_x + \sin{(\xi)}\hat{\sigma}_y$, $\tan{\xi} = x$ and $\tilde{\Omega} = \sqrt{1+x^2}\Omega$. Comparing this expression to the bare qubit Hamiltonian~(\ref{eq:H}), we note that for any given $x$ two differences appear: i) The direction of the adiabatic following on the Bloch sphere is rotated from the ($\hat{\sigma}_x,\hat{\sigma}_z$)-plane to the ($\hat{\sigma}_\xi,\hat{\sigma}_z$)-plane and ii)
the effective Rabi frequency, equivalent to a magnetic field component in the $\xi$-direction, is increased by a factor $\sqrt{1+x^2}$.
These two observations explain the rotated trajectories seen in Fig.~\ref{F1}(a) and we can understand how they influence the integrated value of the Berry curvature in $\braket{\hat{p}}$.
The rotation i) implies that the diabatic correction to the trajectory is no longer in the $\hat{\sigma}_y$-direction which is the one recorded by the momentum of the meter. Instead it is in the direction perpendicular to $\hat{\sigma}_\xi$. 
Hence, the value recorded by the meter is reduced by a factor $\cos({\xi}) = 1/\sqrt{1+x^2}$. At the same time, the increased strength ii) facilitates a more adiabatic evolution, yielding an additional reduction in the diabatic correction by a factor $1/\sqrt{1+x^2}$. This can also be understood from the expression for the Chern number \eqref{Ctwo_level}. The Chern number is a topological constant, independent on the specific value of $\Omega_1$, so from the expression~(\ref{eq:Berry_y}) for the Berry curvature we must conclude that the local diabatic correction $\braket{\hat{\sigma}_y}$ scales as $1/\Omega_1$ (now $1/\tilde{\Omega}$).

For any finite value of $x$, we thus expect to underestimate the local Berry curvature by a factor $\frac{1}{1+x^2}$, and hence for our Gaussian position distribution for the meter $|\varphi(x)|^2$ the  Chern number is underestimated  by the factor 
\begin{align}\label{beta}
\beta &= \int_{-\infty}^{\infty} \frac{|\varphi(x)|^2}{1+x^2} \, dx \nonumber
\\ 
&=\sqrt{\frac{\pi}{2(\Delta x)^2}}\mathrm{e}^{\frac{1}{2(\Delta x)^2)}}\mathrm{Erfc}\left[\sqrt{\frac{1}{2(\Delta x)^2}}\right],
\end{align} 
where $\mathrm{Erfc}[\cdot]$ is the complementary error function.
In Fig.~\ref{F1}(d), we observe an almost complete match between this expression and the observed reduction in $\braket{\hat{p}}$ as a function of $\Delta x$.
By simply dividing the inferred Chern number in the left panel of Fig.~\ref{F1}(c) by this factor, we obtain the right panel where all the curves follow the same dependence. This observation encourages measurements  with a meter prepared in an initial state with finite $\Delta x$ and exploitation of the simple scaling factor to infer the Chern number from the change in $\braket{\hat{p}}$,
\begin{align}\label{eq:Ccorr}
C_{\mathrm{estimate}} = \braket{\hat{p}}/\beta.
\end{align}

\subsection{Measurement uncertainty}
\begin{figure*}
\begin{center} 
\includegraphics[scale=0.3]{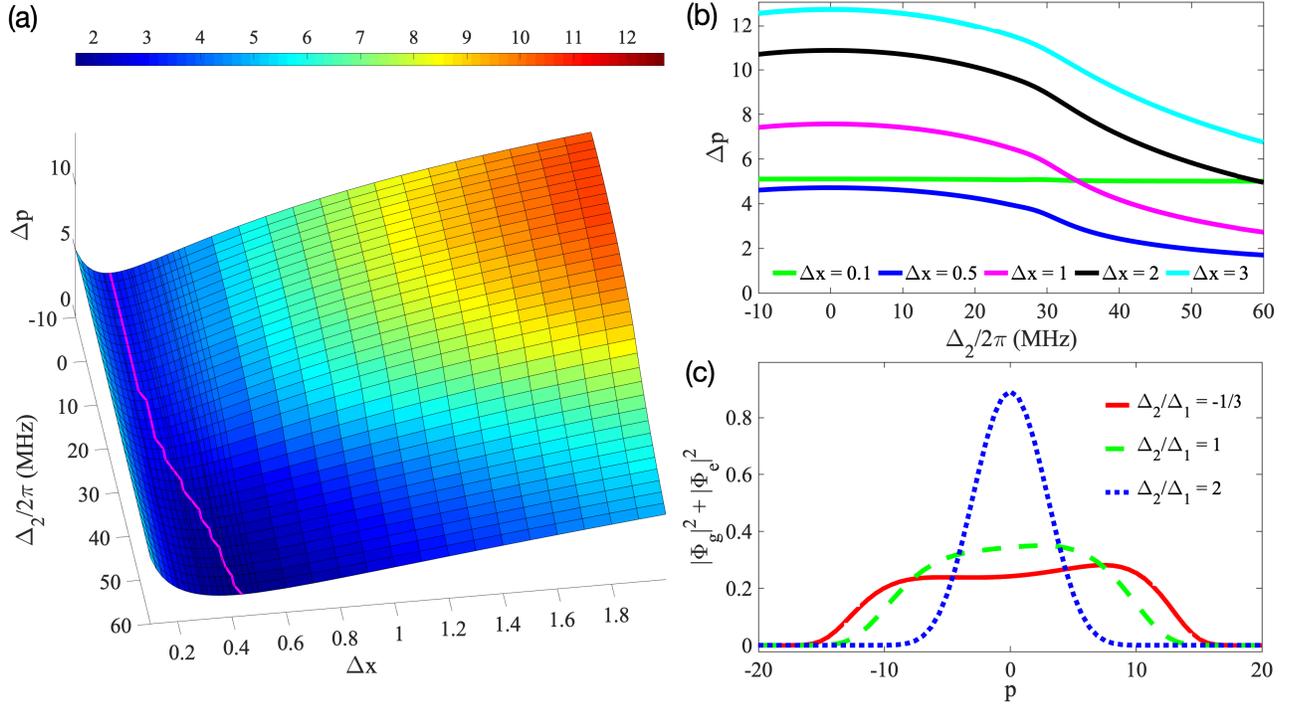}
\end{center}
\caption{Standard deviation $\Delta p$ and $p$-distribution of the meter wave function. 
Results are shown for $t_q = 1\mu s$, $\Delta_1 = 2\pi\times 30$MHz and $\Omega_1 = 2\pi\times 10$MHz.
(a) Surface plot showing the meter momentum standard deviation $\Delta p$ for different values of $\Delta_2$ and $\Delta x$. The magenta curve tracks the values of $\Delta x$ which minimize $\Delta p$ for each value of $\Delta_2$.
(b) The same data as in (a) but shown as curves for $\Delta x \in (0.5, 0.1, 1, 2, 3)$ from below at $\Delta_2 = -2\pi\times10$ MHz.
(c) The momentum probability density $|\Phi_e|^2 + |\Phi_g|^2$, with $\Phi_e$ and $\Phi_g$ defined in \eqref{eq:Phi}, at the final time $t_q$. Results are shown for $\Delta x = 1$ and different values of $\Delta_2/\Delta_1$. 
\label{fig:F2}}
\end{figure*}
We have seen that, contrary to the ideal situation, the integrated value of $\hat{\sigma}_y$
depends critically on the value of $\Delta x$. At the same time, our ability to correctly extract the Chern number by measuring the momentum of the meter, is restricted by the standard deviation $\Delta p$ of the momentum observable in the meter state at the final time $t_q$. 
In the absence of the interaction $\hat{H}_I$, the momentum space wave function, corresponding to the real space wave function~(\ref{eq:fullState}), is itself a Gaussian with a standard deviation given by
$\Delta p = 1/(2\Delta x)$, but the coupling to the qubit may lead to both a change of the mean value $\braket{\hat{p}}$ and the uncertainty $\Delta p$ after the quench. That this indeed occurs is evident in Figs.~\ref{fig:F2}(a)~and~(b). For large $\Delta_2$, the standard deviation $\Delta p$ approaches $1/(2\Delta x)$ for all values of $\Delta x$ because here the qubit is effectively tuned out of resonance with the meter.
For small $\Delta_2$, however, the interaction leads to a deformation and in general a broadening of the momentum distribution.

The magenta line in Fig.~\ref{fig:F2}(a) tracks the value of $\Delta x$ which minimizes the momentum uncertainty $\Delta p$ for different values of $\Delta_2$. The optimal value for $\Delta_2 \simeq 0$ is $\Delta x \simeq 0.23 < 1\sqrt{2}$, corresponding to, in fact, a substantial antisqueezing in the $p$ observable! As $\Delta_2$ increases beyond the transition point ($\Delta_2 = \Delta_1$), the minimum moves to larger values of $\Delta x$.

We can acquire additional insight on this mechanism by studying the momentum wave functions of the meter, conditioned on each of the two qubit states $\ket{g}$ and $\ket{e}$ at the final time $t_q$,
\begin{align}\label{eq:Phi}
\Phi_{\alpha}(p) = \frac{1}{\sqrt{2\pi}}\int \varphi(x)c_{\alpha}(x,t_q)\mathrm{e}^{-i p x}\,dp,
\end{align}
with $\alpha = g,e$.
The corresponding probability densities are shown for $\Delta x = 1$ and different values of $\Delta_2$ in in Fig.~\ref{fig:F2}(b). 
It is evident that for small values of  $|\Delta_2 |$, the momentum distribution is severely broadened and seems to be bifurcating. As $\Delta_2$ increases across $\Delta_1 = 2\pi\times 30$MHz, the momentum wave function~(\ref{eq:Phi}) retains its Gaussian shape during the quench.

To understand this we evoke that, in the adiabatic limit, for a given value of $x$, the qubit state evolves along the instantaneous (adiabatic) eigenstate of the full Hamiltonian~(\ref{eq:Hfull}) but acquires a geometric phase $\gamma_g(x,t)$ as well as a dynamic phase $\gamma_d(x,t)=-\int_0^{t} E_+(x,t')\,dt'$,
\begin{align}\label{eq:phases}
\ket{\chi(x,t)} = \mathrm{e}^{i\gamma_g(x,t)}\mathrm{e}^{i\gamma_d(x,t)}\ket{\chi_{+}(x,t)},
\end{align} 
where  $\ket{\chi_{\pm}(x,t)}$ are the adiabatic eigenstates with
Bloch vector representations
\begin{align}\label{bloch AD}
\vec{R}_{\pm}(x,t)
= \frac{\pm1}{\sqrt{\Omega^2(1+x^2)+\Delta^2}}
\begin{pmatrix}
\Omega \\ \Omega x\\\Delta
\end{pmatrix},
\end{align}
and 
$E_{\pm}(x,t) = \pm\frac{1}{2}\sqrt{\Omega^2(1+x^2)+\Delta^2}$
is the instantaneous eigenenergy in this state. Due to the symmetry of the problem, the dynamic phase is an even function of $x$, such that the shift in momentum arises solely from the geometric phase, $\braket{p(t_q)} = -\int \, |\varphi(x)|^2\frac{\partial \gamma_g(x,t_q)}{\partial x}\, dx$. Notice that the geometric phase $\gamma_g(x) \equiv \gamma_g(x,t_q)$  accumulated at the final time $t_q$ is, in fact, independent of the value of $t_q$ since the Hamiltonian is varied through a closed parameter manifold. 

For small $x$, $E_+(x,t)\simeq \frac{1}{2} \sqrt{\Delta^2 + \Omega^2} + \frac{\Omega^2 x^2}{4 \sqrt{\Delta^2 + \Omega^2}}+ O(x^4)$.
Upon integration in Eq.~(\ref{eq:phases}), the first ($x$-independent) term produces a trivial contribution to the dynamic phase associated with the bare qubit evolution under the Hamiltonian $\hat{H}_q$. The second order term, on the other hand, yields an  $x$-dependent dynamic phase factor $\mathrm{e}^{-i  f x^2 t_q}$, where $f$ is a positive function of the control parameters $\Delta_1$, $\Delta_2$ and $\Omega_1$. For $\Delta_2 = 0$ and $\Omega_1=\Delta_1$, we find, for instance, $f = |\Delta_1|/8$.
Although the phase factor does not alter the probability density $P(x) = |\varphi(x)\mathrm{e}^{-i  f x^2 t_q}|^2$ of the $x$ observable, it imposes antisqueezing of the $p$ observable as seen by a broadening of the complementary wave function
in Fig.~\ref{fig:F2}(b). The bifurcation of the momentum space wave function and the exact $\Delta x$ dependence of the standard deviation seen in Fig.~\ref{fig:F2}(b) stems from higher order terms in the expansion of $E_+(x,t)$, and we verify by numerical evaluation a perfect agreement with the results shown in Fig.~\ref{fig:F2}(b) when we use the exact form of $E_+(x,t)$.
Consequently, for small $\Delta x$, during the interaction, the momentum variance grows from its initial value $[\Delta p(t=0)]^2 = 1/(2\Delta x)^2$ to the value
\begin{align}\label{eq:DpTilde}
[\Delta p(t=t_q)]^2 =  \frac{1+16f^2t_q^2(\Delta x)^4}{(2\Delta x)^2},
\end{align}
as seen in the small $\Delta_2$ regime of Fig.~\ref{fig:F2}(a).

\begin{figure*}
\begin{center} 
\includegraphics [scale=0.31] {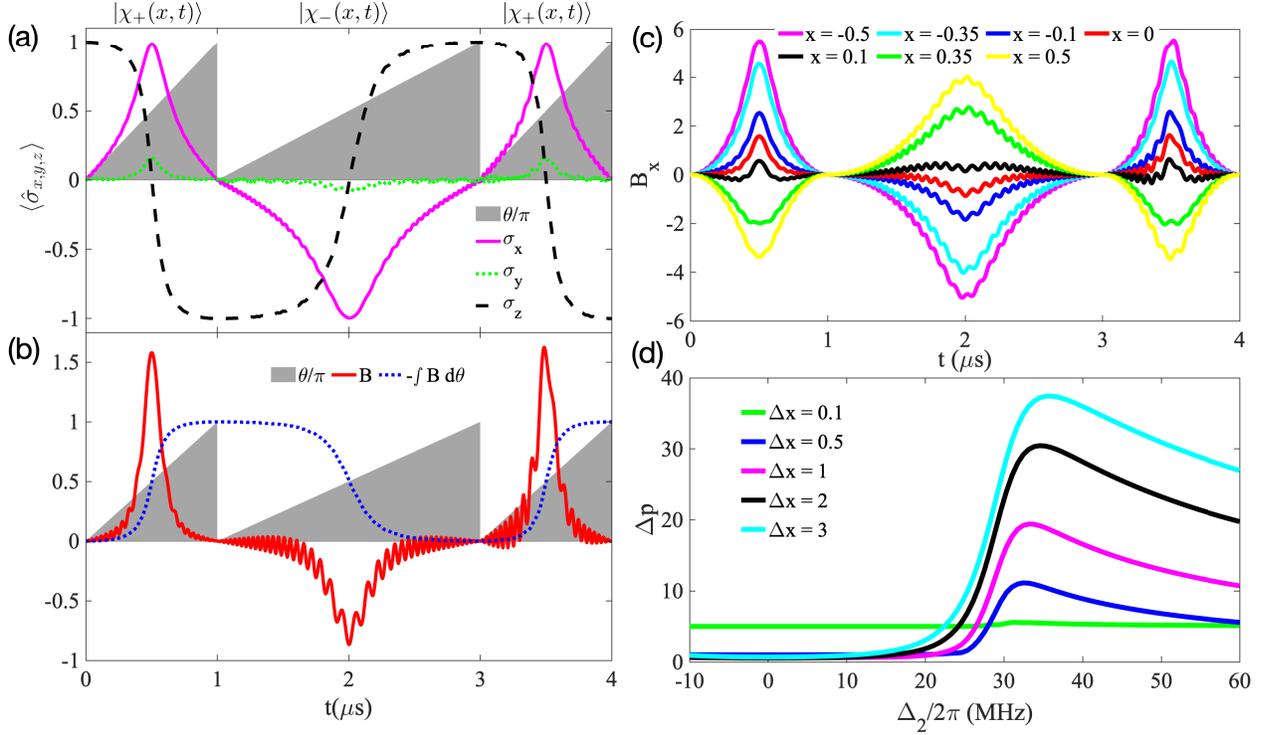}
\end{center}
\caption{Results for the refocusing protocol of Subsection~\ref{sec:refocus}  shown for $t_q = 1\mu s$, $\Delta_1 = 2\pi\times 30$MHz and $\Omega_1 = 2\pi\times 10$MHz.
(a) The shaded gray area represents the change of $\theta/\pi$ during the three quenches of total duration $4t_q$. 
the curves show the ensuing evolution of the qubit Bloch vector components $\langle \hat{\sigma}_{x,y,z}\rangle$ for $\Delta_2 = 0.01\Delta_1$ in the absence of the meter system.
At each stage, the state evolves (apart from a small non-adiabatic correction) along one of the adiabatic eigenstates $\ket{\chi_\pm(x,t)}$ as indicated above the plot.
(b) The corresponding Berry curvature and its integrated value which yields the Chern number at the final time [see Eq.~(\ref{Ctwo_level})].
(c) The value of Eq.~(\ref{eq:Bx}) as a function of time for different $x \in (-0.5, -0.35, -0.1, 0, 0.1, 0.35, 0.5)$ shown from above (below) at $t = 0.5 \mu$s (2$\mu$s) with $\Delta_2 = 0.01\Delta_1$. (d) Standard deviation of the meter momentum $\Delta p$ in the final state as a function of $\Delta_2$ and for different values of $\Delta x \in (0.1, 0.5, 1, 2, 3)$ from below at $\Delta_2/2\pi = 30$MHz. 
\label{fig:F3}}
\end{figure*}
\subsection{Refocusing the meter wave function}\label{sec:refocus}
Our analysis reveals an intricate interplay: the coupling, which allows the information about the topological Berry phase to be transduced to the measurement meter, at the same time introduces a dynamical phase which broadens the momentum distribution, and thereby deteriorates our ability to read out the momentum with high precision. This renders the choice of initial meter state non trivial, and we see that it is neither optimal to use a maximally squeezed or anti-squeezed wavefunction. For this reason, it is an enticing goal to engineer a protocol which maintains the geometric phase while it effectively cancels the dynamical phase, thus allowing a highly sensitive read-out.

This may be achieved by repeating the quench three times \textit{without} reinitialization of the qubit: once with a duration $t_q$, then with a duration $2t_q$ and finally again with a duration $t_q$. For $\Delta_2<\Delta_1$ this amounts to a $3\pi$ rotation of the qubit as shown along with the change in $\theta$ in Fig.~\ref{fig:F3}(a).
Since the geometric phase factor is topologically protected, it does not depend on the specific quench duration, and we see in Fig.~\ref{fig:F3}(b) that for $x =0$, the positive Berry curvature accumulated during the first quench of duration $t_q$ is indeed cancelled by the negative value of Eq.~(\ref{eq:Berry_y}) accumulated during the quench of duration $2 t_q$, such that after the final quench of duration $t_q$, the integrated Berry curvature yields the Chern number $C = 1-1+1 =1$. For $\Delta_2>\Delta_1$, the Berry curvature integrates to zero during each quench, maintaining a Chern number of zero for the full sequence. In Fig.~\ref{fig:F3}(c) we observe the same behaviour of the Berry curvature for different values of $x$.

Due to the finite quench duration, the transition at $\Delta_2 = \Delta_1$ is not completely sharp as seen in Fig.~\ref{F1}(c) as the system is left in super position states after the quench. We notice that this leads to oscillatory behaviour in the transition regime during the multiple quenches introduced here.

To understand the dynamic phase, we consider first the region $\Delta_2<\Delta_1$. Here the qubit follows during the first quench the adiabatic eigenstate $\ket{\chi_{+}(x,t)}$, accumulating a dynamic phase $-\int_0^{t_q}E_+(x,t)\, dt$. It then acquires a phase $-\int_0^{2t_q}E_-(x,t)\, dt$ as it follows the other adiabatic eigenstate $\ket{\chi_{-}(x,t)}$ during the next quench, and finally again a phase $-\int_0^{2t_q}E_+(x,t)\, dt$ while it follows the state $\ket{\chi_{-}(x,t)}$ during the third quench.
Since $E_-(x,t) = - E_+(x,t)$ and the integrated phases are proportional to the integration time, this effectively cancels the dynamic phase 
$ \gamma_{d}(x,t_q) \simeq - (f-2f+f)x^2t_q = 0$.
The result is evident in Fig.~\ref{fig:F3}(d) where we see that for $\Delta_2<\Delta_1$, unlike Fig.~\ref{fig:F2}(b), the width of the final momentum distribution is no longer enhanced by the coupling to the qubit during the protocol.

For $\Delta_2>\Delta_1$, on the other hand, the state returns to the north pole after each quench, following all the time $\ket{\chi_{+}(x,t)}$ and accumulating thereby an enhanced dynamic phase 
$\gamma_{d}(x,t_q) \simeq - (f+2f+f)x^t_q = - 4f x^2t_q$.
When comparing Fig.~\ref{fig:F3}(d) to the results for the single quench [Fig.~\ref{fig:F2}(b)], the detrimental effect of this enhancement on the momentum uncertainty is clear.

To circumvent this issue, we propose to combine a measurement sequence using the standard single quench procedure with one using the triple quench procedure introduced in this subsection. The former would benefit from the low noise in the large $\Delta_2$ regime [see Fig.~\ref{fig:F2}(b)] and the latter would produce data with very small uncertainty in the $\Delta_2<\Delta_1$ regime as seen in Fig.~\ref{fig:F3}(d).

Finally, it is interesting to note that with the triple quench protocol, the almost discontinuous jump in standard deviation around the transition point ($\Delta_2 = \Delta_1$) qualifies the variance in $p$ itself as a good indicator of the distinct topological phases. For large $\Delta x$, the experimentalist would observe almost no noise for $\Delta_2<\Delta_1$ and huge signal fluctuations for $\Delta_2 > \Delta_1$.

\section{Conclusion and Outlook}\label{sec:Conclusion}
In conclusion, we have proposed a new scheme to measure the Chern number and thus characterize a topological transition in a qubit state manifold. Rather than directly measuring the qubit, we propose to introduce a measurement meter which effectively integrates the Berry curvature in its (generalized) momentum observable and thereby allows the Chern number to be read out at the final time. The backaction from the coupling to the meter changes the evolution of the qubit which affects the value of the Chern number such that it is no longer an integer. However, we can analytically understand this mechanism which allows us to introduce a simple correction factor in the estimate from the meter measurement.

At the same time, the interaction between the qubit and the meter introduces a dynamical phase which depends on the meter position. This results in an enhancement of the momentum uncertainty and thereby deteriorates our ability to read out the Chern number at the final time. We propose a simple protocol, relying on three separate quenches, which overcomes this issue by effectively cancelling the dynamical phases while retaining the geometrical phase evolution.

We presented the procedure in terms of a single spin-1/2 system, but in an experimental implementation it may be beneficial to employ a large ensemble of non-interacting spins, coupled to the meter as
$\hat{H}_I = -g(t)\sum_{i=0}^N \hat{x}\hat{\sigma}^{(i)}_y$. This would directly provide a $\sqrt{N}$ enhancement of the signal-to-noise-ratio, and it is an intriguing possibility to investigate if spin squeezing \cite{MaJian2011, Giovannetti2011} can provide a further \textit{quantum} enhancement 
in a setting like this. Likewise, our choice of a meter prepared in a Gaussian state may not be optimal, and there may be improvements to gain by considering alternative initializations such as Fock states or Schrödinger cat states \cite{GiovannettiPRL2006}.

Finally, we want to emphasize that the ability to obtain the time integrated value of a system observable is not restricted to non-adiabatic corrections. The coupling to the meter may alternatively incorporate time dependent factors designed to accumulate, e.g.,  specific frequency-components of perturbations on the system. Our method to separately address the contributions to mean values and to variances of the meter observables may hence also find applications in general metrology challenges.

\section*{Acknowledgements}

This work was financially supported by the Young fund of Jiangsu Natural Science Foundation of China (Grant No. BK20180750). P. X was also supported by the Scientific Research Foundation of Nanjing University of Posts and Telecommunications (NY218097) and the National Natural Science Foundation of China under Grant No. 11847050. S. L. Z was supported by the Key-Area Research and Development Program of GuangDong Province (Grant No. 2019B030330001) and  the National Natural Science Foundation of  China (Grant No. 91636218).
A.~H.~K and K.~M. acknowledge support from the Villum Foundation and the European Union FETFLAG program, Grant No. 820391 (SQUARE).


\begin{thebibliography}{60}


\bibitem{Kosterlitz1973} J. M. Kosterlitz, and D. J. Thouless, Ordering, metastability and phase transitions in two-dimensional systems, J. Phys. C {\bf6}, 1181 (1973).
\bibitem{Thouless1982} D. J. Thouless, M. Kohmoto, M. P. Nightingale, and M. den Nijs, Quantized Hall Conductance in a Two-Dimensional Periodic Potential, Phys. Rev. Lett. {\bf49}, 405 (1982).
\bibitem{Haldane1983} F. D. M. Haldane, Nonlinear Field Theory of Large-Spin Heisen- berg Antiferromagnets: Semiclassically Quantized Solitons of the One-Dimensional Easy-Axis Neel State, Phys. Rev. Lett. {\bf50}, 1153 (1983).


\bibitem{Klitzing1980} K. v. Klitzing, G. Dorda, and M. Pepper, New method for high-accuracy determination of the fine-structure constant based on quantized Hall resistance. Phys. Rev. Lett. {\bf45}, 494 (1980).
\bibitem{Tsui1982} D. C. Tsui, H. L. Stormer, and A. C. Gossard, Two-dimensional magnetotransport in the extreme quantum limit. Phys. Rev. Lett. {\bf48}, 1559 (1982).
\bibitem{Bernevig2006} B. A. Bernevig, T. L. Hughes, and S.-C. Zhang, Quantum spin Hall effect and topological phase transition in HgTe quantum wells. Science {\bf314}, 1757 (2006).
\bibitem{Hasan2010} M. Z. Hasan and C. L. Kane, Colloquium: topological insulators. Rev. Mod. Phys. {\bf82}, 3045 (2010).
\bibitem{Moore2010} J. E. Moore, The birth of topological insulators. Nature {\bf464}, 194 (2010).

\bibitem{Carusotto2019} T. Ozawa, H. M. Price, A. Amo, N. Goldman, M. Hafezi, L. Lu, M. C. Rechtsman, D. Schuster, J. Simon, O. Zilberberg, and I. Carusotto, Topological photonics. Rev. Mod. Phys. {\bf91}, 015006 (2019).
\bibitem{Dalibard2019} N. R. Cooper, J. Dalibard, and I.B. Spielman, Topological bands for ultracold atoms. Rev. Mod. Phys. {\bf91}, 015005 (2019).
 

\bibitem{DanWei2018} D.-W. Zhang, Y.-Q. Zhu, Y. X. Zhao, H. Yan, and S.-L. Zhu,
Topological quantum matter with cold atoms, Adv. Phys. {\bf67}, 253 (2018).


\bibitem{Berry1985} M. V. Berry, Classical adiabatic angles and quantal adiabatic phase, J. Phys. A {\bf18}, 15 (1985).


\bibitem{Zhang2005} Y. Zhang, Y.-W. Tan, H. L. Stormer, and P. Kim, Experimental observation of the quantum Hall effect and Berry's phase in graphene, Nature (London) {\bf438}, 201 (2005).


\bibitem{Klitzing1986} K. von Klitzing, The quantized Hall effect, Rev. Mod. Phys. {\bf58}, 519 (1986).

\bibitem{Haldane1988} F. D. M. Haldane, Model for a Quantum Hall Effect without Landau Levels: Condensed-Matter Realization of the "Parity Anomaly", Phys. Rev. Lett. {\bf61}, 2015 (1988).

\bibitem{Nayak2008} C. Nayak, S. H. Simon, A. Stern, M. Freedman, and S. D. Sarma, Non-Abelian anyons and topological quantum computation, Rev. Mod. Phys. {\bf80}, 1083 (2008).

\bibitem{Gritsev2012} V. Gritsev and A. Polkovnikov, Dynamical quantum Hall effect in the parameter space, Proc. Natl. Acad. Sci. USA {\bf109}, 6457 (2012).

\bibitem{Schroer2014} M. D. Schroer, M. H. Kolodrubetz, W. F. Kindel, M. Sandberg, J. Gao, M. R. Vissers, D. P. Pappas, A. Polkovnikov, and K. W. Lehnert, Measuring a Topological Transition in an Artificial Spin-1/2 System, Phys. Rev. Lett. {\bf113}, 050402 (2014).

\bibitem{Roushan2014} P. Roushan, C. Neill, Y. Chen, M. Kolodrubetz, C. Quintana, N. Leung, M. Fang, R. Barends, B. Campbell, Z. Chen, B. Chiaro, A. Dunsworth, E. Jeffrey, J. Kelly, A. Megrant, J. Mutus, P. J. J. O'Malley, D. Sank, A. Vainsencher, J. Wenner, T. White, A. Polkovnikov, A. N. Cleland, and J. M. Martinis, Observation of topological transitions in interacting quantum circuits, Nature (London) {\bf515}, 241 (2014).


\bibitem{Xu2017} P. Xu, A. Holm Kiilerich, R. Blattmann, Y. Yu, S.-L. Zhu, and K. M{\o}lmer, Measurement of the topological Chern number by continuous probing of a qubit subject to a slowly varying Hamiltonian, Phys. Rev. A {\bf96}, 010101(R) (2017).


\bibitem{Gerlach1922} W. Gerlach and O. Stern,  Der experimentelle Nachweis der
Richtungsquantelung im Magnetfeld, Z. Phys. {\bf8},
110 (1922).

\bibitem{Meekhof1996} D. M. Meekhof, C. Monroe, B. E. King, W. M. Itano, and D. J. Wineland, Generation of Nonclassical Motional States of a Trapped Atom Phys. Rev. Lett. {\bf76}, 1796 (1996).


\bibitem{MaJian2011} J. Ma, X. Wang, C. Sun, and F. Nori, Quantum spin squeezing, Phys. Rep. {\bf509}, 89 (2011).

\bibitem{Giovannetti2011} V. Giovannetti, S. Lloyd, and L. Maccone, Advances in Quantum Metrology, Nat. Photonics {\bf5}, 222 (2011).

\bibitem{GiovannettiPRL2006} V. Giovannetti, S. Lloyd, and L. Maccone, Quantum Metrology,
Phys. Rev. Lett. {\bf96}, 010401 (2006).


\end{thebibliography}
\end{document}